\documentclass[11pt,3p,preprint]{elsarticle}
\usepackage[utf8]{inputenc}
\usepackage[T1]{fontenc}
\usepackage{mathtools,booktabs,amsmath,bm,amssymb,natbib}
\usepackage{subfigure,caption}
\numberwithin{equation}{section}

\usepackage{hyperref}

\title{Constraints on axion-like particles with the Perseus Galaxy Cluster with MAGIC}

\author[a]{\textbf{MAGIC Collaboration}: H.~Abe}
\author[a]{S.~Abe}
\author[b]{J.~Abhir}
\author[c]{V.~A.~Acciari}
\author[d]{I.~Agudo}
\author[e]{T.~Aniello}
\author[f,ad]{S.~Ansoldi}
\author[e]{L.~A.~Antonelli}
\author[g]{A.~Arbet Engels}
\author[h]{C.~Arcaro}
% ...

\address[a]{Japanese MAGIC Group: Institute for Cosmic Ray Research (ICRR), The University of Tokyo, Kashiwa, 277-8582 Chiba, Japan}
\address[b]{ETH Z\"urich, CH-8093 Z\"urich, Switzerland}
\address[c]{Instituto de Astrof\'isica de Canarias and Dpto. de  Astrof\'isica, Universidad de La Laguna, E-38200, La Laguna, Tenerife, Spain}
\address[d]{Instituto de Astrof\'isica de Andaluc\'ia-CSIC, Glorieta de la Astronom\'ia s/n, 18008, Granada, Spain}
\address[e]{National Institute for Astrophysics (INAF), I-00136 Rome, Italy}
\address[f]{Universit\`a di Udine and INFN Trieste, I-33100 Udine, Italy}
\address[g]{Max-Planck-Institut f\"ur Physik, D-80805 M\"unchen, Germany}
\address[h]{Universit\`a di Padova and INFN, I-35131 Padova, Italy}
% ...
\address[ad]{also at International Center for Relativistic Astrophysics (ICRA), Rome, Italy}
% ...

\cortext[correspondence]{Corresponding authors: I.Batkovi\'c,  G.D'Amico.  Email: contact.magic@mpp.mpg.de}

\date{\today}
\begin{document}

\begin{abstract}
Axion-like particles (ALPs) are pseudo-Nambu-Goldstone bosons that emerge in various theories beyond the standard model. These particles can interact with high-energy photons in external magnetic fields, influencing the observed gamma-ray spectrum. This study analyzes 41.3~hrs of observational data from the Perseus Galaxy Cluster collected with the MAGIC telescopes. We focused on the spectra the radio galaxy in the center of the cluster: NGC~1275. By modeling the magnetic field surrounding this target, we searched for spectral indications of ALP presence. Despite finding no statistical evidence of ALP signatures, we were able to exclude ALP models in the sub-micro electronvolt range. Our analysis improved upon previous work by calculating the full likelihood and statistical coverage for all considered models across the parameter space. Consequently, we achieved the most stringent limits to date for ALP masses around 50 neV, with cross sections down to $g_{a\gamma} = 3 \times 10^{-12}$ GeV$^{-1}$.
\end{abstract}

\begin{keyword}
Axion \sep  Axion-Like particles \sep Gamma Rays \sep Galaxy Cluster \sep Imaging Atmospheric Cherenkov Telescopes
\end{keyword}

\maketitle


\section{\label{sec:intro}Introduction}

Axions are pseudo-Nambu-Goldstone bosons that emerge after the spontaneous breaking at a large energy scale $f_a$ of a $U(1)$ symmetry, called Peccei-Quinn, originally introduced as a solution to the so-called Strong-CP problem by \citet{peccei:1977} and further discussed in~\citep{Weinberg:1977ma, Wilczek:1977pj}. The original Peccei-Quinn axion had a mass proportional to $f_a$ at the eV scale (visible axion) and was soon experimentally discarded~\citep{1981PhLB..107..159A}. However, it was realized that axion-like particles (ALPs), similar to axions but lighter in mass and having a mass independent on the coupling, arise in many theories beyond the Standard Model, from  four-dimensional extensions of the Standard Model~\citep{Turok:1995ai}, to compactified Kaluza–Klein theories~\citep{Chang:1999si} and especially string theories~\cite{Witten:1984dg,Svrcek:2006yi,Conlon:2006tq}, see e.g., \citet{Jaeckel:2010ni} for a review. These ALPs are natural candidates to constitute the dark matter (DM) in the Universe~\citep{Arias:2012JCAP,ADMX:2020PhRvL}.
The parameter space of ALPs is wide, with reasonable masses from peV to MeV and a couplings below $10^{-10}~\mathrm{GeV}^{-1}$, and for such reason they are also called Weakly Interacting \textit{Slender} Particles (WISPs), as opposed to the more massive Weakly Interacting Massive Particles (WIMPs) at the GeV scale.

ALPs display a coupling to photons, which happens through a two-photon vertex in the presence of the external electromagnetic field expressed as~\citep{Sikivie:1983ip, Raffelt:1987im}:
\begin{equation}
{\mathcal{L}_{{a}_{\gamma\gamma}}}=-\frac{{g}_{{a\gamma}}}{4}{F}_{\mu\nu}{\tilde{F}^{\mu\nu}}a={g}_{{a\gamma}}\vec{E}\cdot\vec{B}\,a,
\label{eq:interaction_hunt_ALPs}
\end{equation}
where $a$ is the ALP field, $g_{a\gamma}$ is the interaction strength, inversely proportional to the Peccei-Quinn symmetry breaking scale $f_a$, and $F_{\mu\nu}$ is the electromagnetic tensor field. {$\vec{E}$ is the electric field of a beam photon, and~$\vec{B}$ is the external magnetic field. Several experimental approaches utilize in-lab strong magnetic fields, such as the ``light-shining-through-a-wall'' class of experiments, in which a laser is shot through a strong magnetic field, and photons are searched for behind a wall that is opaque to photons, and that can only be crossed by ALPs~\cite{Ehret:2010mh,OSQAR:2015qdv,ALPSII:2022MUPB}. Alternatively, the conversion is sought for in resonant cavities, named haloscopes, filled with strong magnetic fields tuned to frequencies where detection of microwave photons converted from invisible axions is possible. The mass of these axions is of the order of $\mu$eV, in some cases probing the conventional models of invisible axions, as well as the case in which axions are viable candidates for the DM~\citep{ADMX:2020PhRvL,CAST-CAPP:2022NatCo}. The interior of the Sun is also supposed to host significant photon-ALP conversions with an ample ALP flux toward the Earth that can be sampled with experiments searching for back-conversion of these ALPs into photons in strong magnetic fields~\citep{Vogel:2015yka,CAST:2017uph}. See \citet{Irastorza:2018dyq,Graham:2015ARNPS} for recent reviews. 

In the following, instead, we make use of the fact that interactions taking place in astrophysical environments influence the high-energy gamma-ray spectrum received at Earth~\citep{DeAngelis:2007wiw,Hooper:2007bq,Horns:2012kw,Galanti:2018upl,Galanti:2022ijh}. The probability of oscillation $P_{\gamma\to a}$  \citep{Sikivie:1983ip,Raffelt:1987im,Hooper:2007bq,Horns:2012kw} depends on the ALP mass $m_a$, the coupling of the ALPs to photons $g_{a\gamma}$, the ambient magnetic field intensity $B$ in the polarization plane of the incoming photon $E_\gamma$, and its coherence length $s$ (also called $B$ domain size)~\citep{Sikivie:1983ip}:
\begin{equation}\label{eq:P_osc}
P_{\gamma\to a}=\sin^2(2\theta)\,\sin^2\left[\frac{g_{a\gamma}\,Bs}{2}\sqrt{1+\frac{E_{crit}}{E_{\gamma}}}\right],
\end{equation}
where $\theta$ is an effective mixing angle, connected to the geometry between the incoming photon and $B$. To mark the reference energy above which the interaction (mixing) of photon beam and ALPs becomes significant and enters the strong-mixing regime, one can define a critical energy parameter $E_{crit}$ that can be expressed as
\begin{equation}\label{eq:E_crit}
E_{crit}\simeq 247\;\left(\frac{m}{\mu\mathrm{eV}}\right)^2\left(\frac{10\:\mu\mathrm{G}}{B}\right)\left(\frac{10^{-11}\:\mathrm{GeV}^{-1}}{g_{a\gamma}}\right)~\mathrm{TeV}
\end{equation}
where $m^2=|m^2_a-\omega^2_{pl}|$ is the difference between the 
ALP mass $m_a$ and the local electron plasma frequency $\omega_{pl}=\sqrt{4\pi\alpha n_e/m_e}$ where $n_e$ and $m_e$ are the electron density and mass. For a value of magnetic field around $\mu$G, coupling at about $10^{-11}\mathrm{GeV}^{-1}$ and ALP masses in the sub-$\mu$eV scale (and assuming a negligible $\omega_{pl}$ which is often the case), $E_{crit}$ is at the GeV-TeV energy scale, and therefore sub-$\mu$eV ALP signatures have been predicted to be observable by gamma-ray instruments~\citep{DeAngelis:2007wiw,Hooper:2007bq,Sanchez-Conde:2009exi,Horns:2012kw,Galanti:2022ijh} for a decade already.

The equations of motion of the photon-ALP system can be solved using the methodology by~\citet{Raffelt:1987im}. The result of the calculation is an average photon survival probability $P^a_{\gamma\gamma}=P_{\gamma\to a\to\gamma}(\bm{B},m_a,g_{a\gamma})$, defined as the probability that the photon initially emitted from the very-high-energy (VHE) gamma-ray source is converted to an ALP and converted back to a photon. 
One can, generally speaking consider four different regions of $B$ (for an extragalactic target as considered in this work): that one at the emission region where gamma rays are emitted, e.g. in ultra-relativistic jets; a second in the region around the source, as for example the core of galaxy clusters (GC); a third is the Intergalactic Magnetic Field (IGMF) and finally the Milky Way (MW) magnetic field. 
The relative importance of each of the contributions to the overall conversion is debated in the literature~\citep{DeAngelis:2007wiw,Hooper:2007bq,Sanchez-Conde:2009exi,Horns:2012kw, Davies:2020uxn}. We will come back to this when discussing the case of GCs under scrutiny of this work. 
A concurring process affecting the probability of observing a high-energy gamma ray from a distant object is the production of electron-positron pairs in scatterings of high-energy gamma rays off 
UV-optical ambient photons of the Extragalactic Background Light (EBL). The EBL is made up of direct stellar light and light reprocessed by intergalactic dust. The probability that the photon survives the EBL is determined by $P^{\mathrm{EBL}}_{\gamma\gamma}\propto \tau(E_\gamma,z)$, which is related to the optical depth $\tau$, that depends on the photon energy $E_\gamma$ and the source redshift $z$. The astrophysical gamma-ray flux $\Phi_{obs}$ observed at Earth is related to the intrinsic one $\Phi_{int}$ at the emission point by a combination of the two effects:
\begin{equation}
\frac{d\Phi_{obs}}{dE}=\frac{d\Phi_{int}}{dE}\cdot P^{a,\mathrm{EBL}}_{\gamma\gamma}(E_\gamma;a,g_{a\gamma}, B,z)
\end{equation}

where $P^{a,\mathrm{EBL}}_{\gamma\gamma}$ (hereafter $P_{\gamma\gamma}$ for simplicity) combines the probability of EBL absorption and ALP oscillation.
Three regimes can be defined as a function of the critical energy: weak, oscillatory and maximal. In the weak mixing regime, where $E_\gamma \ll E_{crit}$, the conversion probability is small and any ALP signature is negligible. In the case when $E_\gamma \gg E_{crit}$, the mixing is maximal, and the conversion probability becomes energy-independent, resulting in a slow curvature of an observed astrophysical gamma-ray spectrum with a corresponding softening or hardening, according to the specific target under scrutiny~\citep{DeAngelis:2008sk, Hooper:2007bq,Horns:2012kw}. The reason is that at the TeV scale, due to strong EBL absorption, if $P_{\gamma\to a}$ is large at the source, it is possible~\citep{Horns:2012kw} that the ALP flux from a faraway target is much larger than its expected photon flux. Even the conversion of a fraction of these ALPs \textit{back} to photons, e.g. in the MW magnetic field, would result in a spectral hardening. However, the exact computation of this softening/hardening requires accurate modeling of both the EBL and the intrinsic flux~\citep{Sanchez-Conde:2009exi,Horns:2012kw}. The situation is different for $E_\gamma\simeq E_{crit}$, where the mixing is oscillatory and this results in the formation of spectral irregularities, or ``wiggles'', in the gamma-ray spectra. One of the first studies of ALPs in the VHE regime was carried out by the H.E.S.S. collaboration, estimating the irregularities induced by the  mixing in the spectrum of the BL~Lac object PKS~2155-304~\citep{HESS:2013udx}. The results of this study have set the coupling value ${g}_{a\gamma} $ to be smaller than $2.1\times{10}^{-11}~\textrm{GeV}^{-1}$ for masses of the ALPs ${m}_{a}$ in the range $(15-60)$~neV~\citep{HESS:2013udx}. The search for these (although tiny) spectral wiggles does not require an accurate knowledge of the intrinsic source flux or the EBL for detection in case of low-redshift objects, as we will outline in this study.

In this work we search for imprints of ALPs in the observed spectrum an active galactic nuclei (AGNs) located in the center of the Perseus GC. Perseus is the brightest X-ray GC, displaying a dense population of electrons and a strong magnetic field at its core~\cite{Churazov:2003hr,Taylor:2006ta}. In its center, Perseus hosts a very bright TeV-emitting radio~galaxy: NGC~1275~\citep{MAGIC:2011wkl,MAGIC:2013mvu, MAGIC:2016jpr,MAGIC:2018xky}. NGC~1275 has been extensively sampled by MAGIC, producing a wealth of scientific results because of its intense flaring activities. Further studies on the energy density in the Perseus cluster and on dark matter can be found in~\cite{MAGIC:2009tyk, MAGIC:2011vay} and~\cite{MAGIC:2018tuz}, respectively. Apart from the sizable MAGIC dataset, Perseus deems to be an interesting target for ALPs searches due to the strong magnetic field permeating the cluster over large distances (in order of hundreds of kpc), as well as for its proximity to Earth which allows to minimize the discrepancies that arise from a different choice of the EBL model. 

A second bright head-tail radio galaxy, IC~310, is located at 0.6~deg off-center and has shown strong flaring activities observed with MAGIC~\citep{MAGIC:2010kda}. The projected angular distance corresponds to about 750~kpc from the GC center. The true distance is probably much larger considering the largest redshift of IC~310, estimated to be $z=0.0189$, in comparison to the redshift of NGC~1275 of $z=0.0176$. Even at its projected distance, the magnetic field appears to be reduced for about a factor 10 (see Fig.~\ref{fig:magnetic_field}), while at its true distance could be much smaller or vanishing. The IC~310 dataset consists of 1.9~h taken on the November 13th, 2012 and it provided a detection of a strong fast flare with a sensitivity of $18$ standard deviation off the residual background, globally less than that of NGC~1275. Considering the turbulent nature of the GC magnetic field, the $P^{a,\mathrm{EBL}}_{\gamma\gamma}$ for NGC~1275 and IC~310 should not strongly differ due to the different location only, but it would be affected by the magnetic field intensity as well. Before modeling the magnetic field in IC~310, we tried a naive combination of the two dataset \textit{assuming the same $P^{a,\mathrm{EBL}}_{\gamma\gamma}$ for both targets.} We found that IC~310 data are only minorly affecting the constraints obtained with NGC~1275 only. We therefore decided not to consider IC~310 altogether.
For the calculation of the photon-ALP oscillation probability, we model the propagation using the~\texttt{gammaALPs} open-source code\footnote{Hosted on GitHub~(\url{https://github.com/me-manu/gammaALPs}) and archived on Zenodo~\citep{meyer_manuel_2021_6344566}.  See~\citep{Meyer:2021pbp} for an overview.},  which also includes the effects of the EBL and the modeling of magnetic fields.

\bigskip
In this work our main interest is to investigate the possible oscillations in the spectra around the critical energy causing spectral anomalies. We describe the signal model in Sec.~\ref{sec:signal} together with the modeling of the magnetic fields. In Sec.~\ref{sec:methodology}, we outline the novel statistical approach used in the analysis. 
Finding no significant spectral anomalies, in Sec.~\ref{sec:results} we compute 99\% CL upper limits on the photon-ALP coupling as a function of the ALP mass. The results are discussed in Sec.~\ref{sec:discussion}. In the Appendices we discuss the systematics of the analysis and provide further validation of the statistical approach.

\section{\label{sec:signal}Data Preparation and Signal Modeling}

Our search for ALP signatures relies on the modeling of the observed high-energy gamma-ray spectrum of NGC~1275 from MAGIC data, and the conversion probability $P_{\gamma\gamma}$. The latter depends on the modeling of the magnetic field at the Perseus Cluster, the IGMF and magnetic field in the Milky Way. These are hereafter described.

\begin{table*}[h!t]
\centering
\Huge
\resizebox{.99\hsize}{!}{
\begin{tabular}{c|lcccc|ccccc}
\hline\hline
Target & Date & Duration & \(N_{\mathrm{on}}\) & $N_{\mathrm{off}}$ & $N_{\mathrm{exc}}$ &$\mathcal{S}$ & Spectrum & \(\Gamma\) & \(\Phi_0/10^{-10}\) & \(E_k\)\\
       &      &    [h]   &                 &                &     &          &            &    &[cm$^{-2}$\;s$^{-1}$\;TeV$^{-1}$] & [TeV]\\
\hline
NGC~1275 &  1 Jan 2017 & 2.5 & 6632 & 6703 & 4397 & 61.3  & EPWL & $-2.31\pm 0.06$ & $12.2\pm 1.0$ & $0.72\pm0.11$ \\
&  02-03 Jan 2017 & 2.8 & 4376 & 6060 & 2356 & 37.8 &  EPWL & $-1.79\pm0.14$ & $11.4\pm2.1$ & $0.29\pm0.04$ \\
 &  Sep 2016 - Feb 2017  & 36.0 & 28830 & 68943 & 5849 & 31.8 & EPWL & \(-2.54\pm0.13\) & \(1.1\pm0.2\) & $0.5\pm 0.12$\\

\hline\hline
Sum &  & 41.3 & 39838 &  81706 & 12602 & 60.8 & -- & -- & -- & --\\
\hline
\end{tabular}
}
\caption{The three datasets used for the analysis. For each dataset we report the observation date, the duration in hours, the global number of events in the ON and OFF region ($N_{\mathrm{on}},N_{\mathrm{off}}$ respectively), number of excess events ($N_{\mathrm{exc}}$), and the significance of the excess signal in the dataset $\mathcal{S}$ (Eq.~\ref{eq:s_lima}). We report the spectral features corresponding to the null hypothesis (no ALP), namely  ``EPWL'' for exponential cut-off power law, including the photon index $\Gamma$, the normalization flux $\Phi_0$ computed at a normalization energy $E_0=0.3$~TeV in all cases, and the cut-off energy $E_k$.}
\label{tab:data}
\end{table*}

\subsection{\label{sec:datasets_preparation}Preparation of the NGC~1275 Dataset}

NGC~1275 is an AGN classified as a radio galaxy, located at the center of the Perseus Galaxy Cluster at the redshift $z=0.0176$. Observations of NGC~1275 with the MAGIC telescopes include about $400~\mathrm{hrs}$ of data over many years~\citep{MAGIC:2009tyk,MAGIC:2010kda,MAGIC:2011wkl,MAGIC:2011vay,MAGIC:2013mvu, MAGIC:2016jpr,MAGIC:2018xky,MAGIC:2018tuz}. For this study we selected the NGC~1275  data from the period of September 2016 to February 2017, corresponding to the period with highest flux from the source. This is motivated by the fact that the spectral distortion introduced by ALPs is small and only observable when the spectral points are very significant, as it is the case during the high states of the source. The NGC~1275 data are further classified into three datasets, including the strong flare activity detected by MAGIC in Jan 2017, the post-flaring state in the same period, and the baseline emission over two consecutive years (see Tab.~\ref{tab:data}). The whole dataset of NGC~1275 includes $\sim 41~\mathrm{hrs}$ of data~\citep{MAGIC:2018xky}.
The data were processed with the proprietary MAGIC Analysis and Reconstruction Software \texttt{MARS}~\citep{Zanin:2013oib}, following the already published analysis~\citep{MAGIC:2010kda,MAGIC:2011wkl,MAGIC:2018xky}. We have converted the so-called MAGIC proprietary \texttt{melibea} files\footnote{\texttt{melibea} files contain reconstructed stereo events information such as estimated energy, direction, and a classification parameter called \textit{hadroness} $h$ related to the likelihood of being a gamma-like event ($h\to0$ for gamma-like candidates).} into the so-called DL3 format. DL3 (Data Level 3) is the standard format adopted by the next-generation Cherenkov Telescope Array (CTA) consortium~\cite{CTAConsortium:2017dvg} as described by \citet{Nigro:2021xcr}. This was motivated by the fact that DL3 data are analyzable with the cross-platform, multi-instrument, \texttt{gammapy}\footnote{\texttt{gammapy} is an open-source \texttt{python} package for gamma-ray astronomy \url{https://gammapy.org/}. 
It is used as core library for the Science Analysis tools of CTA and is already widely used in the analysis of existing gamma-ray instruments, such as H.E.S.S., MAGIC, VERITAS and HAWC.} open-source software~\citep{gammapy:2017}. 
\begin{figure}[h!t]
    \centering
    \includegraphics[scale=0.45]{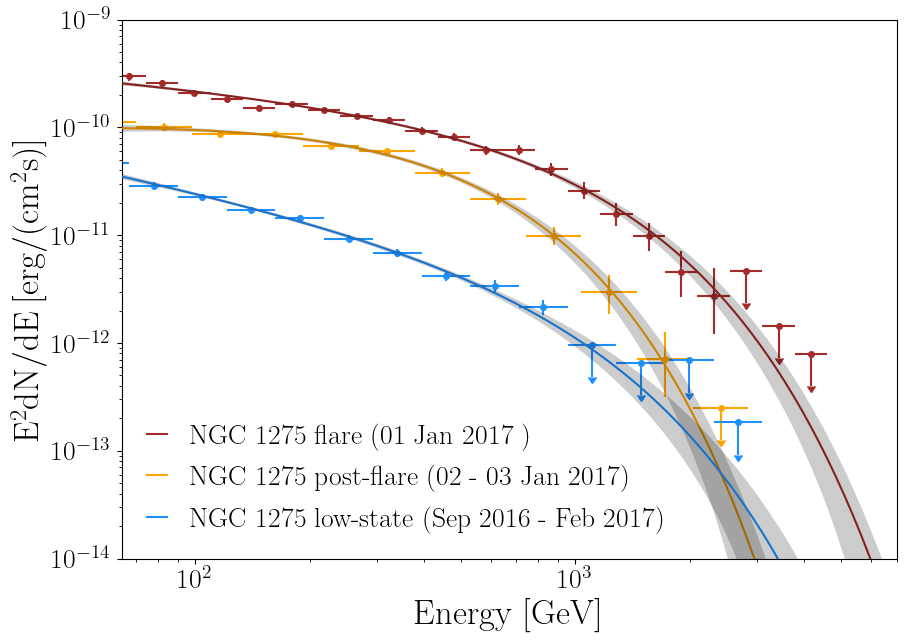}
    \caption{SED of NGC~1275 (different states) obtained with \texttt{gammapy} for the three brightness periods in consideration.}
    \label{fig:SED_Perseus}
\end{figure}

\paragraph{Modeling of NGC~1275 Intrinsic Spectra}
We first present the spectral energy distribution (SEDs) of the three datasets at hand in Fig.~\ref{fig:SED_Perseus}. In the figure, the solid lines represent the best fit of the 
spectral points assuming no--ALP (null hypothesis) and the shaded areas represent the statistical uncertainties on the best fit curve. The best fit curves for the intrinsic energy spectrum, in agreement with Refs.~\cite{MAGIC:2010kda,MAGIC:2018xky} are modeled as a power law with an exponential cut-off (EPWL):
\begin{equation}\label{eq:epwl}
\Phi^i_{int}(E')= \Phi^i_{0} \left(\frac{E'}{E_0}\right)^{\Gamma^i} e^{E'/E^i_{k}},
\end{equation}
for each $i-$th dataset of NGC~1275, where $E'$ is the reconstructed energy, $\Phi_0$ is the normalization flux computed at the energy scale $E_0$. $\Gamma_i$ is the photon index and $E_k$ is the cutoff energy for the EPWL reported in Tab.~\ref{tab:data}.
One can clearly see that NGC~1275 displays spectral variation in function of the source state. 

\medskip
\paragraph{EBL absorption}
A high-energy gamma ray interacts with two main diffuse ambient radiation fields during its propagation through the Intergalactic Medium: the Cosmic Microwave Background (CMB) in the mm range and the UV-optical-IR photons $(10^{-2}-10^{4}$~$\mu$m) of the EBL. If the interaction is efficient, such gamma-ray radiation at high energies is lost through the process of pair production. The UV-optical EBL photon field is the result of the optical-IR direct star light around 1~$\mu$m and the light reprocessed into 100~$\mu$m-range IR light by surrounding dust throughout the evolution of the Universe. This interaction is particularly strong for TeV photons, with optical depths of $\tau\:(z=0.5, E_\gamma=1~\mathrm{TeV})\simeq 4$ and $\tau\:(z=0.5, E_\gamma=10~\mathrm{TeV})\simeq 30$~\citep[][Fig.~12]{Franceschini:2017iwq}.
In this work, the target is in relative proximity  with $z\simeq 0.0176$. As a result, the EBL absorption only plays a minor role at this distance, with an optical depth of $\tau~(300~\mathrm{GeV})\simeq 0.03$ and $\tau~(10~\mathrm{TeV})\simeq 0.4$~\citep[][Fig.~12]{Franceschini:2017iwq}. We model the optical depth due to EBL following \citet{Dominguez:2010bv}. However, there are several other well-motivated models in the literature such as the aforementioned \citet{Franceschini:2017iwq}. There are uncertainties around the true value of the EBL, however, during the past decade, models have been converging to a higher level of agreement. \citet{Stanev:1997zz,Protheroe:2000hp,DeAngelis:2007dqd} realized that the observation of TeV photons was implying an EBL intensity lower than previously expected. This fact first motivated the introduction of the ALP as a way to escape or soften this tension~\citep{Protheroe:2000hp,DeAngelis:2007dqd,Mirizzi:2007hr, Hooper:2007bq,DeAngelis:2008sk,Sanchez-Conde:2009exi,DeAngelis:2011idD,Tavecchio:2012um, Galanti:2015rda}. For this work, the specific choice of the model of \citet{Franceschini:2017iwq} does not have a sizeable impact on the ALP limits, as discussed also by \citet{CTA:2020hii}.

\medskip
\paragraph{Data Binning and Significance}

We have divided the $i-$th dataset in $k-$energy bins both in the ON and OFF regions. The ON region is the Region Of Interest (ROI) in which the signal is expected. Events from the ON region are comprised of both signal and irreducible signal-like background events\footnote{Background events include mostly proton induced events, followed by electrons and heavier cosmic ray nuclei. Trigger and data reconstruction system allow to reject more than $99$\% of the background but an irreducible number of counts usually remains.}. To estimate this number of signal events we use three background control OFF regions in which no signal is expected. The signal is then estimated by the number of \textit{excess} (EXC) events over the estimated number of background events in the ON region, and normalized with an acceptance parameter $\alpha$ between ON and OFF observation. In Tab.~\ref{tab:data} we report the total number $N_{\mathrm{on}},N_{\mathrm{off}},N_{\mathrm{exc}}$ events for the three datasets, as well as the significance $\mathcal{S}$ of $N_{\mathrm{exc}}$, computed both for the individual datasets and a joined one, following Eq.~27 of \citet{Li:1983fv}:
\begin{equation}\label{eq:s_lima}
\mathcal{S}=\sqrt{2\left[N_{\mathrm{on}}\ln\left(\dfrac{(\alpha+1)N_{\mathrm{on}}}{\alpha(N_{\mathrm{on}}+N_{\mathrm{off}})}\right)+N_{\mathrm{off}}\ln\left(\dfrac{(\alpha+1)N_{\mathrm{off}}}{N_{\mathrm{on}}+N_{\mathrm{off}}}\right)\right]}
\end{equation}

\subsection{\label{sec:signal_modeling}Modeling of ALP induced signal}

The presence of ALPs represents our \textit{alternative hypothesis}. According to Eq.~(\ref{eq:E_crit}), we are sensitive in the sub-$\mu$eV, so we prepare a scan of a parameter space with 154 models of ALPs, logarithmically spaced between $4\:\times10^{-9}$~eV and $1\:\times10^{-6}$~eV in mass $m_a$, and $5\:\times10^{-13}\mathrm{~GeV}^{-1}$ and $5\:\times10^{-10}\mathrm{~GeV}^{-1}$ in coupling $g_{a\gamma}$. 
We computed $P_{\gamma\gamma}(E_\gamma;\bm{B})$ using \texttt{gammaALPs} for each of these points, as a function of the different magnetic fields. 

\medskip
\paragraph*{Magnetic fields modeling}

Specific studies for the magnetic field of Perseus $B^S$ are found in \citet{Churazov:2003hr} and \citet{Taylor:2006ta}. A recent comparison between magnetic field models in Perseus was also made by \citet{CTA:Perseus}. Given the large extension of the core and the present magnetic field, the number of domains $N$ crossed by the photon beam is very large and therefore the effective magnetic field encountered $\langle B(r)\rangle=0$, while the RMS can be computed as the average B-field intensity of $\langle B_0\rangle$ following the recipe of \citet{Meyer:2014epa}. Further parameters defined in \texttt{gammaALPs} for the magnetic field of Perseus are taken from \citet{Fermi-LAT:2016nkz}: the electron spatial indices of \citet[][Eq.~4]{Churazov:2003hr} set at $n_0=3.9\cdot 10^{-2}\mathrm{~{cm}^{-3}}$ and density parameter $\beta=1.2$ at 80~kpc, $n_2=4.05\cdot 10^{-3}\mathrm{~{cm}^{-3}}$ and $\beta_2=0.58$ at 280~kpc, the extension of the cluster $r_{Abell}=500$~kpc, and the scaling of the $B$ field with the electron density parameter $\eta=0.5$. The turbulence is modeled in accordance with the A2199 cool-core cluster with maximum and minimum turbulence scale $k_L=0.18~\mathrm{kpc}^{-1}$ and $k_H=9~\mathrm{kpc}^{-1}$ respectively and turbulence spectral index $q=-2.8$  following~\citet{Vacca:2012up}.  These parameters are summarized in Tab.~\ref{tab:magnetic_fields} (upper row). In App.~\ref{app:magnetic_field} we compare our choice of GC magnetic field, based on the work of \citep{Fermi-LAT:2016nkz}, to the recent one used in~\citep{CTA:Perseus}.  

As for the strength of $B^{\mathrm{IGMF}}$, there are still large uncertainties, with upper limits at the nG scale~\citep{Grasso:2000wj} and lower limits at the $10^{-8}$~nG scale~\citep{MAGIC:2022piy}. When inserting such values in Eq.~(\ref{eq:P_osc}) one finds that, at TeV-scale energies, the photon-ALP beam is in the weak-mixing regime, with negligible contributions to the photon-ALP mixing.

Finally, the modeling of $B^{\mathrm{MW}}$ is based on the work of \citet{Jansson:2012pc}. The magnetic field is modeled with a turbulent component, with $10^{-2}$~pc domain size, and a regular component that varies between $1.4-4.4~\mu$G from the Sun vicinities to the exterior.

\section{\label{sec:methodology}Statistical Framework}

The primary objective of the analysis discussed in this article is to evaluate the hypotheses of the existence of signatures of ALPs in the observed gamma-ray spectra. These signatures are derived by setting the coupling constant $g_{a \gamma}$ and mass $m_a$ to the values assumed to occur in nature. The null hypothesis assumes that no--ALP effects are present, implying that only EBL absorption occurs. We achieve this objective by employing a likelihood maximization method.

We define a binned likelihood as follows
\begin{equation}
\mathcal{L}( g_{a \gamma}, m_a, \bm{\mu}, \bm{b} , B | \bm{D} )= \prod_{i,k} \mathcal{L}_{i,k} ( g_{a \gamma}, m_a, \bm{\mu_i} , b_{i,k}, B| \bm{D}_{i,k} ), \label{Eq:Likelihood}
\end{equation}
where $\bm{\mu_i}$ are the SED nuisance parameters (flux amplitude, spectral index and cut-off energy, see table \ref{tab:data}) for the $i$--th sample in our dataset, $b_{i,k}$ are the expected background counts in the OFF region,  and $\bm{D}_{i,k} =(N_{\mathrm{on}}^{i,k}, N_{\mathrm{off}}^{i,k})$ are the  number of ON and OFF events observed in the $k$--th energy bin from the $i$--th sample (see Sec.~\ref{sec:signal}). With $B$ we indicate one possible magnetic-field realization. The likelihood is by definition the probability of observing the data $D_{i,k}$ assuming the model parameters $g_{a \gamma}$ and  $m_a$ to be true:
\begin{equation}
    \mathcal{L}_{i,k} =  \mathcal{P} \left( N_{\mathrm{on}}^{i,k} \; |\;  s_{i,k}+\alpha\,b_{i,k}\right) \times  \mathcal{P}\left(N_{\mathrm{off}}^{i,k} \; | \;   b_{i,k} \right) 
    \label{Eq:PoissonTerm}
\end{equation}
with $\mathcal{P}$ being the Poisson probability mass function for observing $n$ counts with expected count rate $r$: 
$
    \mathcal{P}(n|r) = r^n e^{-r}/n!
$.
The parameter $\alpha$ is the exposure ratio of the ON and OFF region (see Sec.~\ref{sec:signal}), while  $s_{i,k}$ is  the expected signal counts  in the energy bin $\Delta E_k$ in the ON region for the $i$--th sample:
\begin{equation}
    s_{i,k} = \int_{\Delta E_k} \; dE \; \Phi_{obs}^i ( E ; \;  g_{a \gamma}, m_a, \bm{\mu_i}, B,z).
    \label{Eq:EnBin_Integral}
\end{equation}

In Eq.~(\ref{Eq:EnBin_Integral}) we have introduced the observed flux for the $i$--th sample
\begin{equation}
    \Phi_{obs}^i = \int dE'\;\Phi_{int}^i(E';\bm{\mu_i}) P_{\gamma\gamma} (E') \cdot \mathrm{IRF}^i(E | E')   .
    \label{Eq:ConvolutionIRF}
\end{equation}
Thus, in order to perform the integrals in Eq.~(\ref{Eq:ConvolutionIRF}) and Eq.~(\ref{Eq:EnBin_Integral}), and get the likelihood expression from Eq.~(\ref{Eq:PoissonTerm}) and Eq.~(\ref{Eq:Likelihood}), we need to determine the following quantities:
\begin{itemize}
    \item the instrument response function $\mathrm{IRF}^i(E|E')$ for the $i$--th sample, i.e. the probability of detecting an event with true energy $E'$ and assigning it an energy $E$;
    
    \item the survival probability 
  
    $ P_{\gamma\gamma} (E' ;a,g_{a \gamma},B,z)$
    in which both ALPs induced absorptions in GC and MW, together with EBL attenuation in the IGMF, are taken into account.
    
    \item The intrinsic energy spectra $\Phi_{int}^i$ described  in Sec.~\ref{sec:signal} for each dataset. See also Tab.~\ref{tab:data}.
\end{itemize}
We have therefore 9 nuisance parameters $\bm{\mu_i}$ coming from the intrinsic spectrum: 3 for each of the EPWLs of the 3 states of NGC~1275. Further nuisance parameters of the analysis are the magnetic-field realization $B$, as discussed in Sec.~\ref{sec:signal}, and the expected background counts $b_{i,k}$ which are fixed to the values $\hat{b}_{i,k}$ that maximize it for a fixed $s_{i,k}$, as shown by~\citet{Rolke:2004mj}:

\begin{equation}
    \hat{b}_{i,k} = \frac{N + \sqrt{N^2 + 4(1+ 1/\alpha)\; s_{i,k} \;  N_{\mathrm{off}}} }{2(1+ \alpha)},
    \label{Eq:background_profiling}
\end{equation}
with $N \equiv N_{\mathrm{on}}^{i,k} + N_{\mathrm{off}}^{i,k} - (1+ 1/\alpha) \; s_{i,k}$.

Given the likelihood in Eq.~(\ref{Eq:Likelihood}), the statistic $\mathcal{TS}$ is defined as:

\begin{equation}
\begin{split}
\mathcal{TS}(g_{a \gamma}, m_a) &= -2 \Delta \ln \mathcal{L}
\\  
   &= -2 \ln \frac{\mathcal{L}( g_{a \gamma}, m_a,\hat{ \bm{\mu}}, \hat{\bm{b}} , \hat{B}   | \bm{D} )}{ \hat{\mathcal{L}}},
    \label{Eq:Likelihood_ratio}
\end{split}
\end{equation}

where $\hat{\mathcal{L}}$ is the maximum value of the likelihood over the parameter space, while $\hat{\bm{\mu}}$ and $ \hat{B} $ are obtained from profiling the likelihood, i.e. by fixing them to the values that maximize the likelihood for a given coupling $g_{a \gamma}$ and mass $m_a$. 

For the nuisance parameter $B$ instead, given the limitations of computational power, it is improbable that the magnetic-field realization $B$ which maximizes the likelihood function $\mathcal{L}$ is included among the simulated magnetic-field realizations.
Thus, instead of profiling over $B$,  we sort the likelihoods $\mathcal{L}$ in each ALP grid point in terms of the magnetic-field realization. At this point, for each ALP grid point we use the likelihood value that corresponds to a specific quantile $Q=0.95$ of the obtained distribution of $\mathcal{L}.$\footnote{If one could have been sure about the presence of the $B$ field that maximizes $\mathcal{L}$ in the simulations, then a proper treatment of the nuisance parameter $B$ would correspond to putting $Q=1$, i.e.  profiling over $B$.  This procedure for the treatment of the nuisance parameter $B$ is the same adopted in \cite{CTA:2020hii} in which it was found (and confirmed by our analysis) that putting $Q = 0.95$ and not to $1$ is insensitive to the \textit{ad-hoc} choice of number (100 in our analysis) of realizations.}

The statistic defined in Eq.~(\ref{Eq:Likelihood_ratio}) is known as the likelihood ratio. According to the Neyman-Pearson lemma \citep{Neyman:1933}, it is the goodness-of-fit test with maximum \textit{power}, and according to Wilks' theorem \citep{Wilks:1938} it follows a $\chi^2$-distribution with 2  degrees of freedom. This is because the log-likelihood defined in Eq. \ref{Eq:Likelihood_ratio} is a function of only two parameters, $m_a$ and $g_{a \gamma}$.
In our analysis, however, the primary conditions necessary for a direct application of Wilks' theorem are not satisfied. For example, one prerequisite stipulates that two distinct points within the parameter space should yield two unique predictions. Unfortunately, this condition does not hold up when considering values of the couplings $g_{a\gamma}$ close to zero (i.e., there is no ALP effect). In such cases, any variation in the mass $m_{a}$ will inevitably lead to identical predictions, thus violating this essential criterion.
Therefore assuming a $\chi^2$-distribution with two degrees of freedom for  the statistic $\mathcal{TS}(g_{a \gamma}, m_a)$ would lead to a wrong coverage. For this reason, we have computed the correct coverage by getting the effective distribution of the statistic from Monte Carlo (MC) simulations. 

In previous works~\citep{CTA:2020hii} this was done by computing these distributions for few ALP points (generally 2 or 3 points that produce the most pronounced features in the energy flux) and taking the most conservative one, i.e. the one with larger 0.95 (or 0.99) quantile. This was motivated also by the computing power needed to extract these distributions for different points. In our analysis we applied a more accurate approach that consists of computing the distribution of the statistic $\mathcal{TS}(g_{a \gamma}, m_a)$ for each of the 154 points in the ALP parameter space. In this way, we can now directly translate a certain  $\mathcal{TS}(g_{a \gamma}, m_a)$ into a significance for excluding the ALP hypothesis $(g_{a \gamma}, m_a)$, expressed in standard deviation of the corresponding Gaussian or the $z-$score. The resulting exclusion significance for each ALP hypothesis considered in this analysis is discussed in  Sec.~\ref{sec:results}.

\section{\label{sec:results}Results}

Using the datasets of Tab.~\ref{tab:data} and following the prescription described in detail in  Sec.~\ref{sec:methodology}, we compute the statistic $\mathcal{TS}(g_{a\gamma}, m_a) $ in Eq.~(\ref{Eq:Likelihood_ratio}) for each of the 154 points in our ALP parameter space. As described in further details in App.~\ref{app:coverage}, these observed statistics are used to compute the rejection significance of the ALP hypotheses.

\begin{figure}[h!t]
    \centering
    \includegraphics[width=0.55\linewidth]{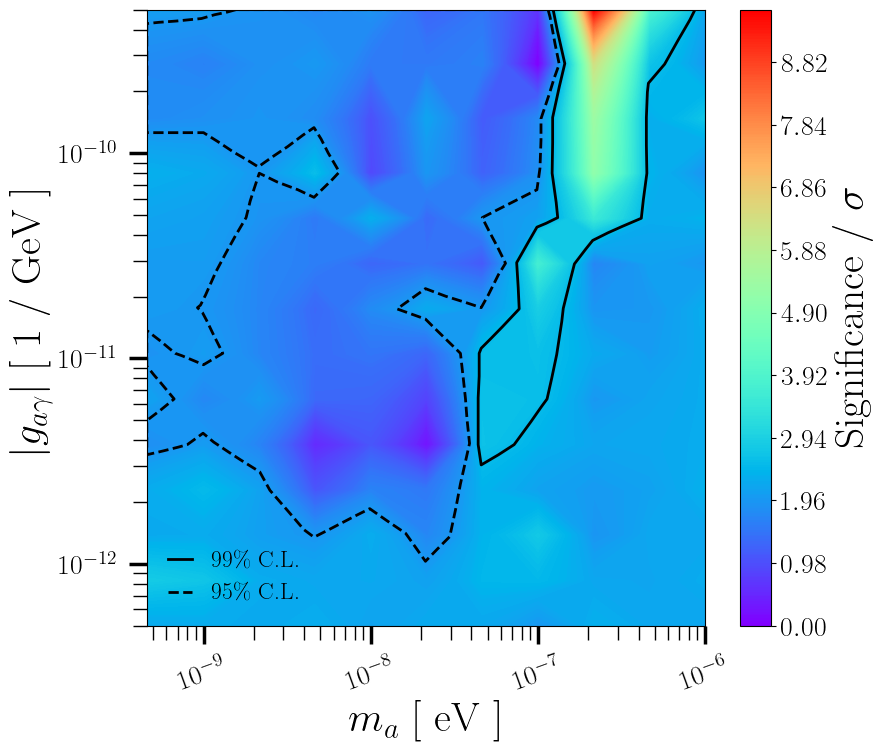}
    \caption{The likelihood-ratio statistic $\mathcal{TS}$ of Eq.~(\ref{Eq:Likelihood_ratio}) is computed over 154 ALP points with  $m_a$ and $g_{a \gamma}$ using the data in Tab.~\ref{tab:data}. For each point, the obtained statistic is then compared to the distribution of $\mathcal{TS}$ one  would get assuming the corresponding ALP hypothesis $m_a$ and $g_{a \gamma}$ to be true. The obtained p-value is converted in the 1-dimensional-Gaussian equivalent standard deviations $\sigma$ (also known as $z-$scores). See App.~\ref{app:coverage} for more details.  The black dashed line  shows a significance of  1.96 $\sigma$ while the black solid one a significance of 2.58  (corresponding to a  95\% and 99\% confidence level, respectively).}
    \label{fig:sigma}
\end{figure}

The rejection significance is shown in Fig.~\ref{fig:sigma} for each point (smoothed for graphical purposes) expressed in numbers of the 1-dimensional-Gaussian equivalent standard deviations $\sigma = \sqrt{2}\,\rm{erf}^{-1} ( \rm{CL} )
$, where $\rm{erf}^{-1}$ is the inverse of the error function and CL is the confidence level for excluding the hypothesis (see App.~\ref{app:coverage} for more details). 
The dark red area corresponds to ALP models that are excluded above 5 standard deviations. Dark blue area corresponds to ALP models that are better in agreement with the data, i.e. they have a low significance rejection.
The model that better agrees with the observation is the one corresponding to $m_a = 1.0 \times 10^{-7}$ eV and  $g_{a \gamma} = 2.71\times10^{-10} ~\textrm{GeV}^{-1}$.  
The null hypothesis of no--ALP effect is disfavored with a $\sim 2 \sigma$ confidence level in favor of the alternative hypothesis, which is not enough to claim any discovery of ALP effects. 
As further discussed in App.~\ref{app:counts},  the spectral points of Fig.~\ref{fig:SED_Perseus} are nicely fit with simple dependency as Eq.~(\ref{eq:epwl}):  the null hypothesis yielded: 

\begin{equation}
-2 \ln \mathcal{L}( g_{a \gamma} = 0 \;\rm{GeV}^{-1}, m_a = 0 \; \rm{ eV},\hat{\bm{\mu}}, \hat{\bm{B}}   | \bm{D} ) = 62.2,
\label{Eq:null_hyp_ts}
\end{equation}
 
which is an expected value considering the total number of degrees of freedom\footnote{The total number of degrees of freedom are given by the difference between the number of energy bins and the number of free parameters used in the model, summed over all datasets. Such a value corresponds for this analysis to 60.}, indicating a good fit to the data. However, the alternative hypothesis corresponding to $m_a = 2.15 \times 10^{-8}$ eV and  $g_{a \gamma} = 3.81 \times 10^{-12}~\textrm{GeV}^{-1}$   demonstrated an even better agreement with

\begin{equation}
-2 \ln \mathcal{L}( g_{a \gamma} , m_a ,\hat{\mu}, \hat{B}   | \bm{D} ) = 55.4.
\label{Eq:best_fit_ts}
\end{equation}

Following Eq.~(\ref{Eq:Likelihood_ratio}) we obtain for the null hypothesis a statistic of $\mathcal{TS} = 6.8$. As discussed in App.~\ref{app:coverage}, assuming the null hypothesis to be true a more extreme value of 6.8 would have been observed  only $4.2 \%$ of the times, which corresponds to a rejection significance for the null hypothesis of $2.03 \,\sigma$. Since the null hypothesis is already excluded at $ 95.8 \%$ CL in favour of the alternative hypothesis, the exclusion region of the ALPs parameter space obtained here will be shown at $ 99 \%$ CL.

\section{\label{sec:discussion}Discussion}

\subsection{\label{sec:point_by_point}Point by point coverage computation.}

 The computation of the rejection significance is done through the likelihood ratio test statistic of Eq.~(\ref{Eq:Likelihood_ratio}), and, as discussed in Sec.~\ref{sec:methodology}, the use of the Wilks'~\citep{Wilks:1938} theorem for the nested hypothesis cannot be blindly applied. For this reason, for each point of the ALP parameter space the correct coverage is  obtained through MC simulations (see App.~\ref{app:coverage}). In our work we have managed to compute the coverage for each point, which allowed us to calculate the $z-$score reported in Fig.~\ref{fig:sigma}. This is a relevant improvement with respect to earlier similar computations such as done in \citet{CTA:2020hii} where it is explicitly mentioned that the coverage of the test statistic is not computed point by point but only for 3 points, among which the one that yields the most conservative exclusion is used. This approach was thereafter needed due to the substantial computational resources required to generate MC simulations for all ALP points.  

\begin{figure*}[h!t]
    \centering
    \includegraphics[width=0.85\linewidth]{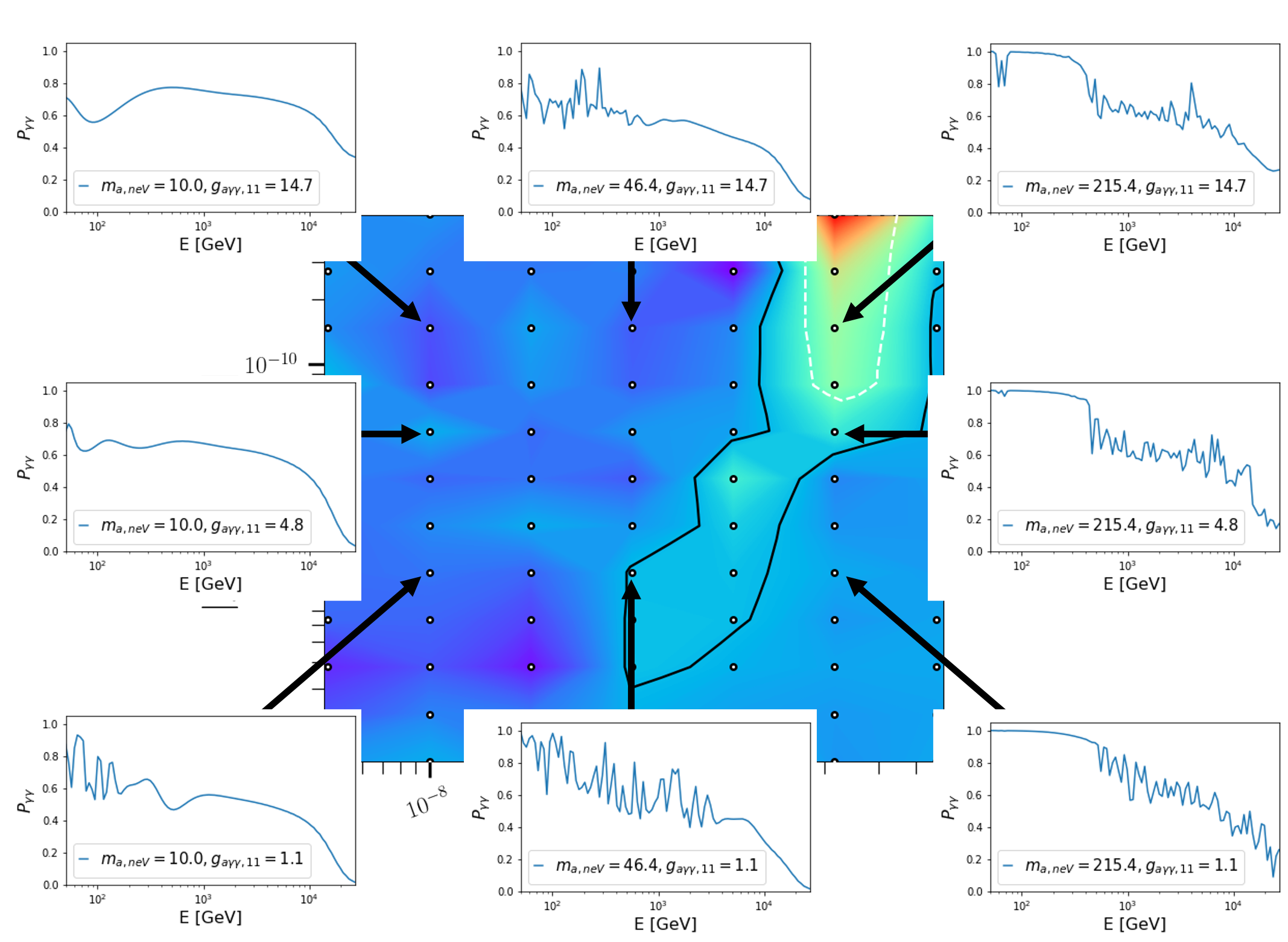}
    \caption{$P^a_{\gamma\gamma}$ for a selection of models in our scan of the parameter space. ALP mass (in~$\textrm{neV}$) and coupling (in~$10^{-11} \textrm{GeV}^{-1}$) are displayed in the inlays. The background image is the significance distribution of Fig.~\ref{fig:sigma} with the addition of the 99\% CL curve (dashed white line) obtained with the conservative coverage computation method of \citet{CTA:2020hii} (see also Sec.~\ref{sec:discussion}). The black dashed line is the 99\% CL curve obtained from the point-by-point coverage computation as described in Sec.~\ref{sec:methodology}. }
    \label{fig:alp_spectra}
\end{figure*}

In Fig.~\ref{fig:alp_spectra} we compare our method with the assumption of \citet{CTA:2020hii}. This is shown in the significance inlay of the figure were, besides our 99\% CL excluded region, we also report the 99\% CL  region that we would have obtained using the previous, more conservative coverage-computation method of \citet{CTA:2020hii}. One can clearly see that the conservative coverage method computation significantly reduces the strength of the limits.

\subsection{\label{wiggles_vs_jumps}Effect of wiggles and jumps}
In Fig.~\ref{fig:alp_spectra} we also report the corresponding $P_{\gamma\gamma}$ for a selection of 8 points in the parameter space. It is interesting to note the evolution of this probability: going from smaller to larger $m_a$, $P_{\gamma\gamma}$in  general becomes more oscillating; going from large to small $g_{a \gamma}$ the oscillations change pattern in an irregular way. 

In the figure we clearly see how the strongest constraints come from a region in which $P_{\gamma\gamma}$ \textit{has sudden jumps rather than just wiggles}: compare e.g. the right column of $P_{\gamma\gamma}$ versus the central one. This follows from the fact that spectral jumps are more easily identified in the observed gamma-ray spectra, or alternatively that wiggles are too small to be detected due to the limited statistic and energy resolution of the instrument. 
This has important consequences in the search for ALP signatures with IACTs considering that in previous publications, the search was focused specifically on wiggles. This is further discussed in the next section.

\subsection{\label{sec:results_extrapolation}Comparison with current limits and CTA projection}

Our limits displayed in Fig.~\ref{fig:sigma} show the highest significance for ALP masses $\sim~m_a=200$~neV for couplings to photons between $g_{a \gamma} = 5.0 \times 10^{-11}~\textrm{GeV}^{-1}$ and $g_{a \gamma} = 5.0 \times 10^{-10}~\textrm{GeV}^{-1}$. However, similar limits obtained with H.E.S.S.~\citep{HESS:2013udx} or forecast with CTA~\citep{CTA:2020hii} are also sensitive to lower ALP masses around 10~neV. We decided to further investigate this discrepancy. In particular, the results from the CTA were obtained by extrapolating a portion of the NGC~1275 dataset that we are using to generate this result:  \citet{CTA:2020hii} consider that during the lifetime of CTA Perseus could be observed for 260~hrs, during which NGC~1275 would be in the baseline emission state for 250~hrs and in flaring state for 10~hrs. The authors model the baseline and flaring state with the values measured by MAGIC and reported here~\citep{MAGIC:2010kda,MAGIC:2018xky}. 

\begin{table}[h!t]
\centering
\small
\resizebox{.70\linewidth}{!}{
\begin{tabular}{c|lccccc}
\hline\hline
Target & State & Duration & $N_{\mathrm{on}}$ & $N_{\mathrm{off}}$ & $N_{\mathrm{exc}}$ & $\mathcal{S}$ \\
&   &  [h]  &   &   & \\
\hline
NGC~1275 & Flare & 10 & 18154 & 12046 & 14138 & 129.0 \\
(mock) &  Baseline  & 252 & 201735 & 482674 & 40852 & 83.9 \\
\hline\hline
& Sum & 262 & 219889 & 494720 & 54990 & 110.0 \\
\hline
\end{tabular}
}
\caption{The two datasets of mock NGC~1275 data used to cast our limits to compare them with \citet{CTA:2020hii}. For each dataset we report the status, the duration in hours, the global numbers $N_{\mathrm{on}}$ and $N_{\mathrm{off}}$ of events in the ON and OFF region, respectively, and the significance of the excess signal in the dataset $\mathcal{S}$. We do not report the spectral parameter for the null hypothesis (no ALP) as they correspond to those in Tab.~\ref{tab:data} for the respective states. }
\label{tab:data_mock}
\end{table}

We therefore adopt the same approach and recompute our limits as if we had taken 250~hrs of baseline and 10~hrs of flaring states. As done in \citep{CTA:2020hii}, we neglect the post-flaring state of NGC~1275, see Tab.~\ref{tab:data_mock}. To do so we are using the previously defined datasets where the observations are convoluted with the IRFs, ultimately giving us the predicted number of counts. To extend our flaring state and baseline to 10~hrs and 252~hrs respectively, we simulated with \texttt{gammapy} $\sim4$ and $\sim7$ times more total predicted counts in comparison to the original datasets of the flaring state and baseline used in the main part of this article.

 \begin{figure}[h!t]
    \centering
    \includegraphics[width=0.55\linewidth]{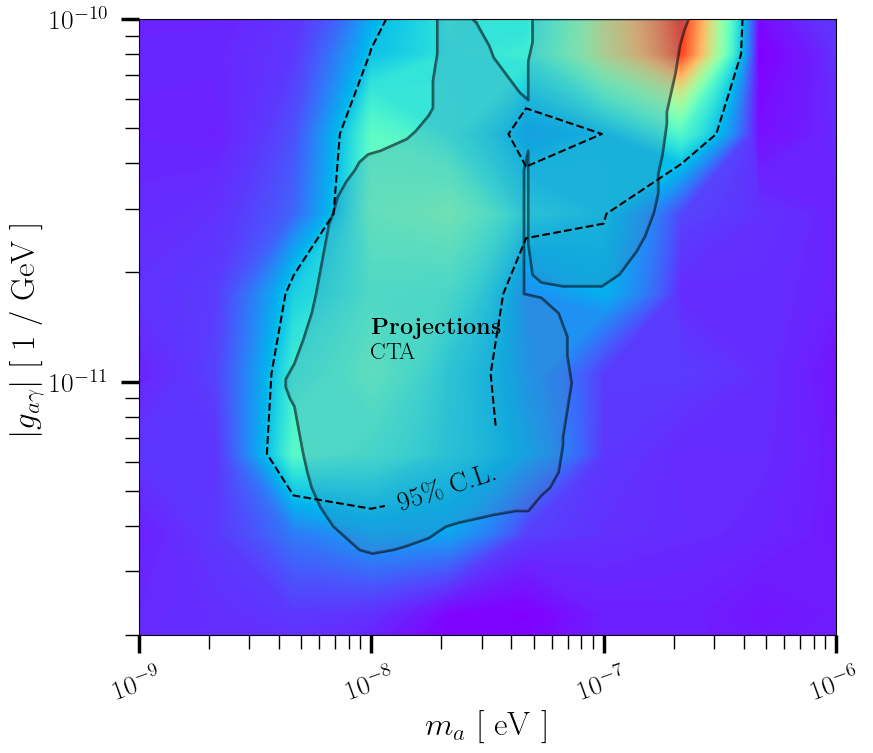}
    \caption{95\% CL exclusion region obtained with CTA projection from \citet{CTA:2020hii} (black line) compared to the projection of the MAGIC limits (dashed black line, 95\% CL) obtained in this work, assuming a larger observation time of 262~hrs corresponding to a global excess significance $\mathcal{S_{\rm{Li\&Ma}}}$ of 110 as in Tab.~\ref{tab:data_mock}. The background color map shows the rejection significance expressed in number of standard deviations $\sigma$. The color coding is the same of Fig.~\ref{fig:sigma}.}
    \label{fig:cast}
\end{figure}

The significance distribution is shown in Fig.~\ref{fig:cast}. We can clearly see that adding significantly more data allows to become sensitive to the parameter region with ALP masses around $1-10$~neV, in agreement with~\citet{CTA:2020hii}. When comparing our findings with those from the Cherenkov Telescope Array (CTA), it is essential to acknowledge that the CTA limits might be more conservative. This is due to their consideration of discrete step-wise variations in the effective area, which have been smoothed at the energy-resolution scale and were assumed to have an amplitude of $\pm 5\%$. These variations were taken to occur at energies where one subsystem of telescopes begins to assume dominance in terms of point-source sensitivity. Therefore, a direct comparison should account for these methodological differences.

Lastly, in Fig.~\ref{fig:context},
we juxtapose the limits established by MAGIC with the currently accessible limits~\citep{HESS:2013udx,Fermi-LAT:2016nkz,Zhang:2018wpc,Cheng:2020bhr,Guo:2020kiq,CTA:2020hii} within the corresponding range of the ALPs parameter space. Our constraints are consistent with limits obtained using similar astrophysical data analysis techniques, and represent the most competitive constraints for ALP masses $m_a$ in the range of $40 - 90 $~neV.

\begin{figure}[h!t]
    \centering
    \includegraphics[width=0.55\linewidth]{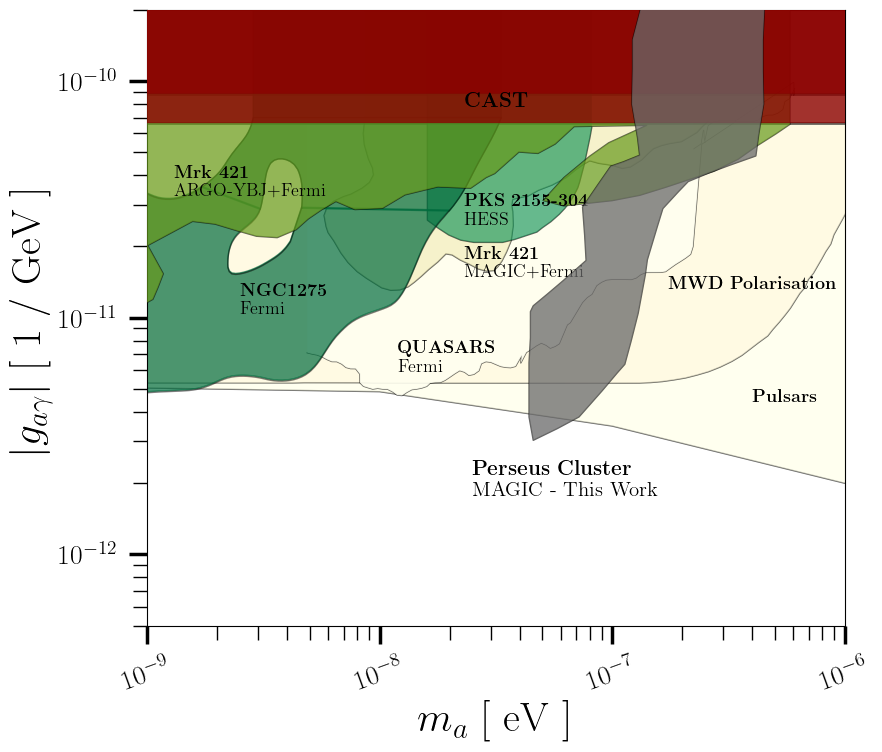}
    \caption{The 99\% CL limits obtained with this work in comparison with current 95\% CL limits in similar part of the parameter space, gathered in~\citep{AxionLimits}.}
    \label{fig:context}
\end{figure}

\section{\label{sec:conclusion}Summary and Conclusions}

In this work we have analyzed 41~hrs of high-energy gamma-ray data coming from the direction of the Perseus galaxy cluster in search for spectral irregularities induced by ALPs in the sub-$\mu$eV mass range. We have used gamma-ray beams of the radio galaxy in the center of the cluster: NGC~1275, during its high emission state to have a significant detection. We have tested the alternative hypothesis (presence of ALP) on 154 points regularly selected in the ALP parameter space. For each model we have computed $P_{\gamma\gamma}$ over 100 realizations of the magnetic field around the target. The test statistic, once calibrated, does not provide significant detection, which allowed us to compute 99\% CL exclusion upper limits in the ALP parameter space. These limits are shown in Fig.~\ref{fig:context} in comparison with other results and constrain ALP masses in the range $40-400$~neV. The excluded area matches that by earlier results and forecast for CTA. In particular in Fig.~\ref{fig:cast} we show how larger observation times or significance of this target would allow to constraint also part of the parameter space at lower masses, around the neV. \\
In Fig.~\ref{fig:alp_spectra} we have computed the significance point by point showing that this allows to improve the constraining power of the data with respect to vigorously conservative assumption on the coverage. In the same figure we have shown how IACTs are sensitive to ALP spectral induced jumps rather than wiggles, a fact which is usually not appreciated. \\
To date, these results offer the strongest constraints on ALP masses in the range of $40-90$~neV, with the greatest sensitivity for ALP masses of $m_a=40$~neV, reaching the photon-axion coupling down to $g_{a \gamma} = 3.0 \times 10^{-12}~\textrm{GeV}^{-1}$.

\section*{CRediT authorship contribution statement}
\textbf{I. Batkovi\'c:} Writing  - original draft. Leading the data and statistical analysis. \textbf{G. D'Amico:} Writing - review \& editing. Leading the statistical analysis and assisted in the interpretation of results. \textbf{M. Doro:} Writing - review \& editing. Supervising and project planning, leading the interpretation of results. Assisting in the data analysis. \textbf{M. Manganaro:} Writing - review \& editing. Assisting in the data analysis and interpretation of results.
\textbf{The MAGIC collaboration:} The rest of the authors have contributed in one or several of the following ways: design, construction, maintenance and operation of the instrument(s); preparation and/or evaluation of the observation proposals; data acquisition, processing, calibration and/or reduction; production of analysis tools and/or related Monte Carlo simulations; discussion and approval of the contents of the draft.

\section*{Declaration of competing interest}
The authors declare that they have no known competing financial interests or personal relationships that could have appeared to influence the work reported in this paper.

\section*{Acknowledgments}
We would like to thank Manuel Meyer for useful discussions. We would also like to thank the Instituto de Astrof\'{\i}sica de Canarias for the excellent working conditions at the Observatorio del Roque de los Muchachos in La Palma. The financial support of the German BMBF, MPG and HGF; the Italian INFN and INAF; the Swiss National Fund SNF; the grants PID2019-104114RB-C31, PID2019-104114RB-C32, PID2019-104114RB-C33, PID2019-105510GB-C31, PID2019-107847RB-C41, PID2019-107847RB-C42, PID2019-107847RB-C44, PID2019-107988GB-C22 funded by the Spanish MCIN/AEI/ 10.13039/501100011033; the Indian Department of Atomic Energy; the Japanese ICRR, the University of Tokyo, JSPS, and MEXT; the Bulgarian Ministry of Education and Science, National RI Roadmap Project DO1-400/18.12.2020 and the Academy of Finland grant nr. 320045 is gratefully acknowledged. This work was also been supported by Centros de Excelencia ``Severo Ochoa'' y Unidades ``Mar\'{\i}a de Maeztu'' program of the Spanish MCIN/AEI/ 10.13039/501100011033 (SEV-2016-0588, CEX2019-000920-S, CEX2019-000918-M, CEX2021-001131-S, MDM-2015-0509-18-2) and by the CERCA institution of the Generalitat de Catalunya; by the Croatian Science Foundation (HrZZ) Project IP-2022-10-4595 and the University of Rijeka Project uniri-prirod-18-48; by the Deutsche Forschungsgemeinschaft (SFB1491 and SFB876); the Polish Ministry Of Education and Science grant No. 2021/WK/08; and by the Brazilian MCTIC, CNPq and FAPERJ.
I.B. and M.D. acknowledge funding from Italian Ministry of Education, University and Research (MIUR) through the ``Dipartimenti di eccellenza'' project Science of the Universe. G.D'A’s work on this project was supported by the Research Council of Norway, project number 301718.


\newpage
\appendix

\section{Systematics Discussion}
\label{app:systematics}

\subsection{Relevance of magnetic field modeling}
\label{app:magnetic_field}
As discussed in Sec.~\ref{sec:signal_modeling}, the modeling of the magnetic field in Perseus is still only fairly known up to date. To address this, the CTA Consortium recently conducted a detailed study comparing various magnetic field models available for Perseus~\citet[see][Fig.~1]{CTA:Perseus}. For their study, they adopted a configuration based on~\citet{Taylor:2006ta} with a reference magnetic field value $B_0=25\:\mu$G and $\eta=2/3$. All remaining parameters of this modeling are reported in  Tab.~\ref{tab:magnetic_fields} and compared with our primary choice, based on~\citet{Fermi-LAT:2016nkz}.

\begin{table}[h!t]
\centering
\begin{small}
 \begin{tabular} {c|cccccc}
 \hline\hline
 &$B_0$ & $\eta$ & $n_0$ & $n_2$ & $r_{\textrm{core}}/r_{\textrm{core}_2}$ & $\beta/\beta_2$\\
 &$\mu \textrm{G}$ &&$\textrm{cm}^{-3}$&$\textrm{cm}^{-3}$&kpc&\\
\hline
$B$ &10&0.5&$39\cdot10^{-3}$ & $4.05\cdot10^{-3}$ &80~/~280&1.2~/~0.58\\
\hline
$B_{alt}$ &25&$2/3$ & $46\cdot10^{-3}$ & $3.60\cdot10^{-3}$ &57~/~278&1.2~/~0.71\\
\hline
\end{tabular}
\end{small}
\caption{The parameters used for the modeling of the Perseus magnetic field. $B$ is representing the parameters used in the main analysis of this article~\citep{Fermi-LAT:2016nkz}. $B_{alt}$ are taken from~\citet{CTA:Perseus}.}
    \label{tab:magnetic_fields}
\end{table}

\begin{figure}[h!t]
    \centering
    \includegraphics[width=0.45\linewidth]{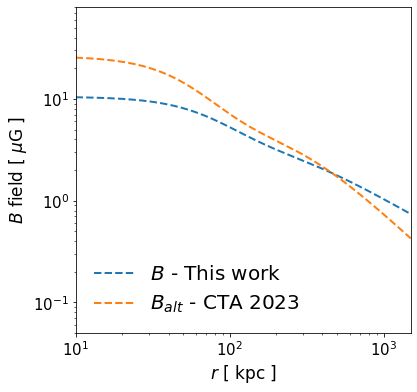}
    \caption{Comparison of magnetic field's radial profile. Blue dashed line is the reference model used in this work. Dashed orange line is the magnetic field model of~\citet{CTA:Perseus}.}
    \label{fig:magnetic_field}
\end{figure}

A visual comparison of the $B(r)$ is reported in Fig.~\ref{fig:magnetic_field}. One can see that \citep{CTA:Perseus} displays a larger magnetic field toward the center of the cluster without models in the region beyond 100~kpc. The effect on the choice of magnetic field on the upper limits is significant and is reported in Fig.~\ref{fig:B_sys_exclusions}.

\subsection{Relevance of the energy scale}
\label{app:energy}
    The MAGIC telescopes reconstruct the energy with a precision of the order of $10-15$\% depending on the energy, which is considered during data reconstruction and an irreducible energy bias, which introduce energy scale uncertainties estimated to be around $\pm15 \%$  ~\citep{ALEKSIC201676}.

    To evaluate this effect, we artificially scaled the ALP energy-dependent signatures in the spectra by $\pm 15\%$ and checked the effects on the bounds. The resulting discrepancies in the exclusion regions are shown in Fig.~\ref{fig:energy_sys}.
    The effect is not negligible, but it does not alter our main conclusions. This uncertainty will be strongly reduced with upcoming IACT arrays, like CTA, whose energy scale systematics are expected to go down to $\sim 4\%$ ~\citep{Gaug_2019}.
    
\begin{figure}[h!t]
    \centering
    \includegraphics[width=0.55\linewidth]{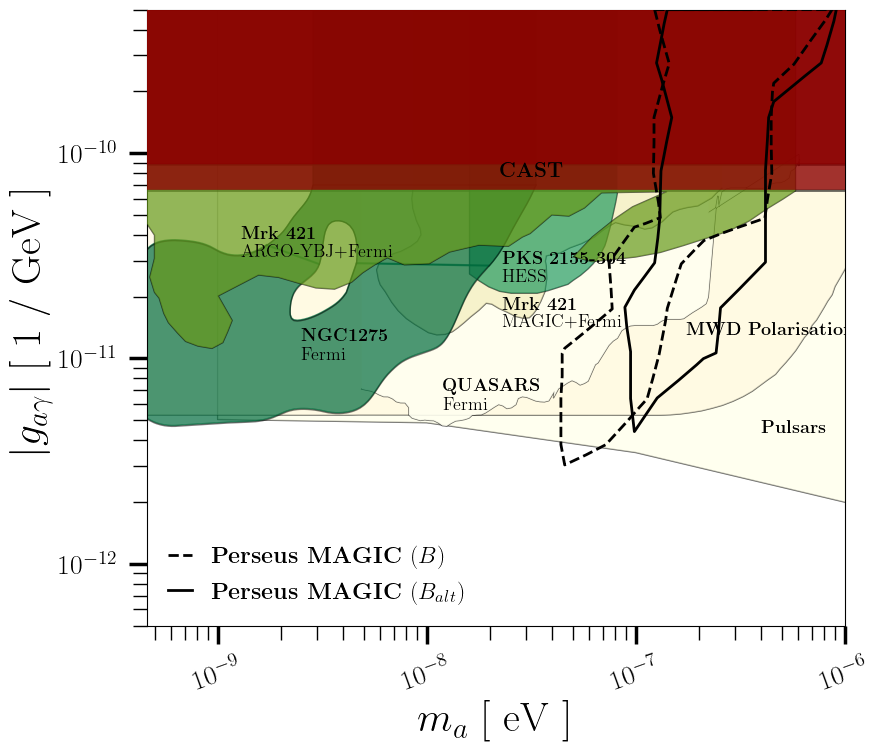}
    \caption{Comparison of the limits in the ALPs parameter space obtained with the Perseus cluster magnetic field from the main part of the article with an alternative magnetic field model used in~\citet{CTA:Perseus}.}
    \label{fig:B_sys_exclusions}
\end{figure}

\begin{figure}[h!t]
    \centering
    \includegraphics[width=0.55\linewidth]{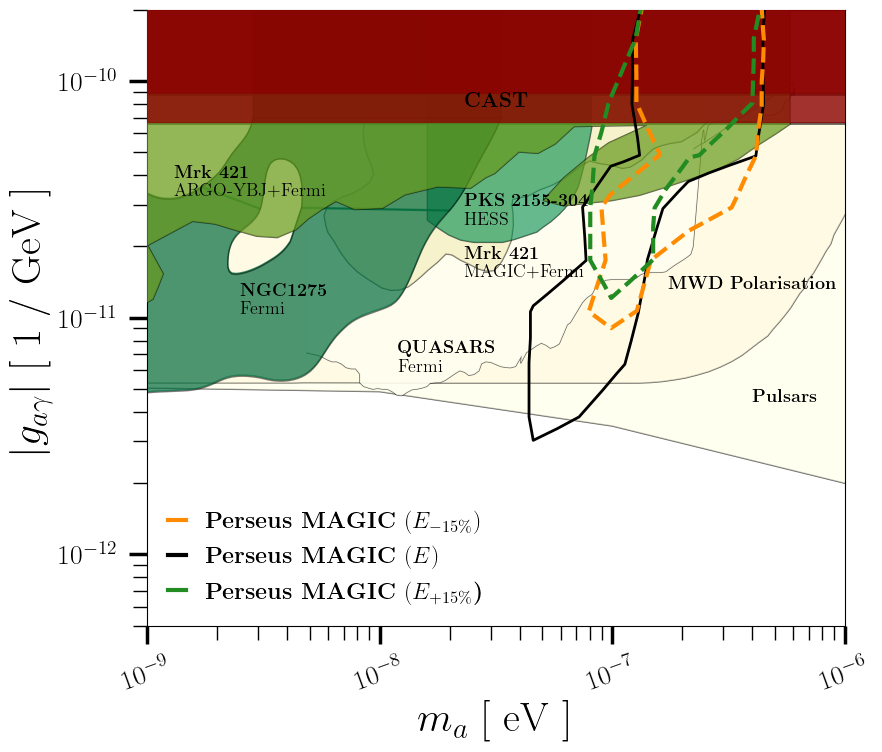}
    \caption{Discrepancies in the exclusion regions resulting from shifting the energy scale by $- 15\%$ (dashed green) and $+ 15\%$ (dashed orange)  in the ALP signatures in the spectra. For comparative purposes, we also depict in dashed black the exclusion regions obtained in this study, as presented in the main text in Fig.~\ref{fig:context}.  }
    \label{fig:energy_sys}
\end{figure}

\section{Computation of the coverage}
 \label{app:coverage}

 The likelihood ratio statistic, as described in (Eq.~\ref{Eq:Likelihood_ratio}), is expected to follow a chi-squared distribution with a number of degrees of freedom equal to the number of independent parameters, according to Wilks' theorem \citep{Wilks:1938}. In our case, there are two independent parameters: the ALP mass ($m_a$) and the axion-photon coupling ($g_{a \gamma}$). However, Wilks' theorem is not applicable for this analysis, necessitating the determination of proper coverage through Monte Carlo simulations. We perform this assessment on a point-by-point basis, in contrast to the approach taken by \citet{CTA:2020hii}, where the most conservative point among the few investigated was selected. 
 In Fig.~\ref{fig:cdf}, we present the Cumulative Distribution Functions (CDFs) of the statistic $\mathcal{TS}(m_a, g_{a \gamma})$ obtained from MC simulations, considering various axion masses ($m_a$) and two distinct axion-photon couplings: $g_{a \gamma} = 5.0 \times 10^{-13}~\textrm{GeV}^{-1}$ (upper plot) and $g_{a \gamma} = 1.7 \times 10^{-11}~\textrm{GeV}^{-1}$ (bottom plot). It is noteworthy that, for the lowest coupling considered in this analysis ($g_{a \gamma} = 5.0 \times 10^{-13}~\textrm{GeV}^{-1})$, the CDFs of the statistic exhibit minimal variation across different $m_a$ values. This observation is consistent with expectations, as the ALP effects on the observed SED are relatively subtle for such a low coupling value, leading to only minor changes in the statistic's distribution when the ALP mass is altered. 

 \begin{figure}[h!b]
    \centering
\includegraphics[width=0.45\textwidth]{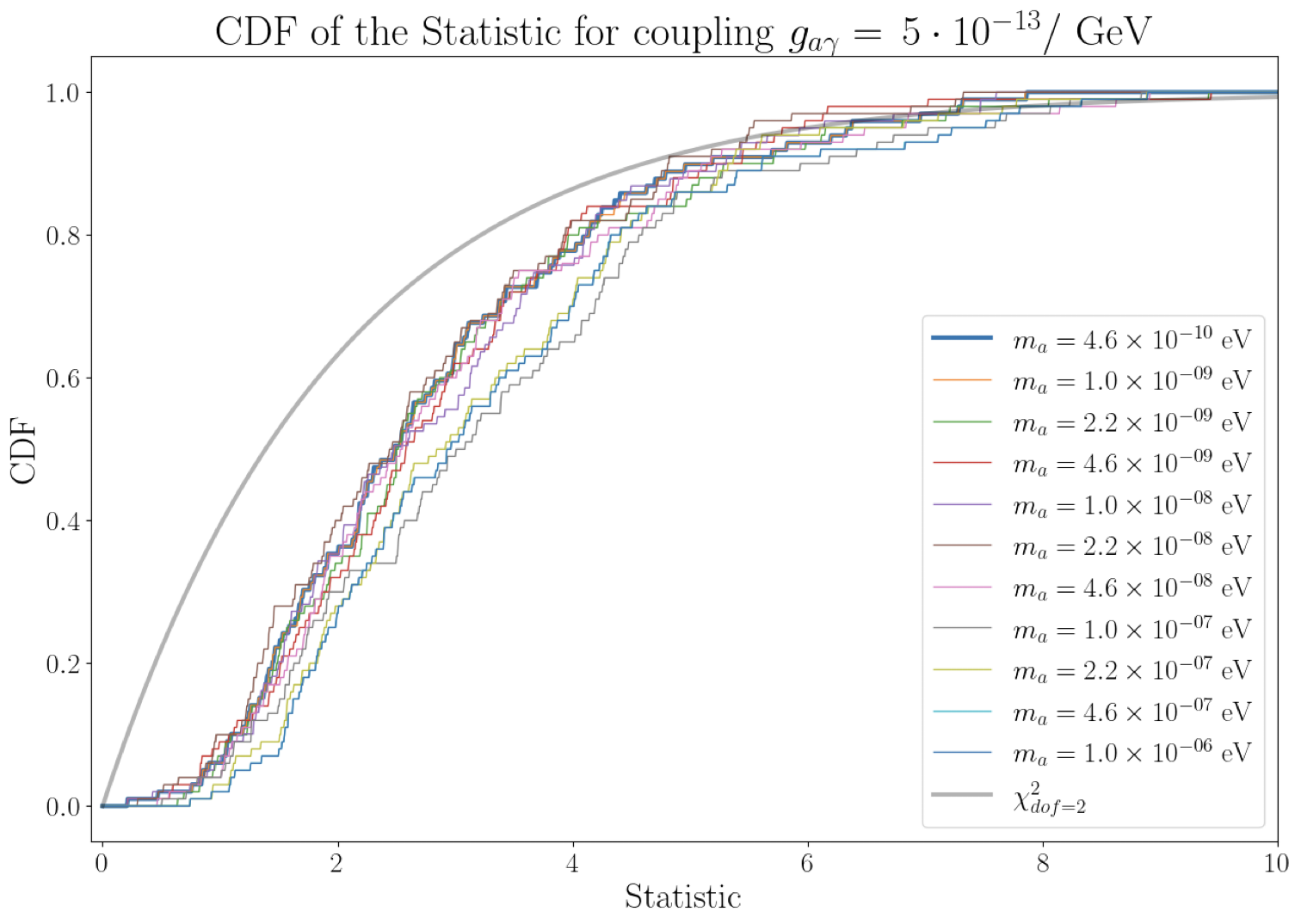}
\includegraphics[width=0.45\textwidth]{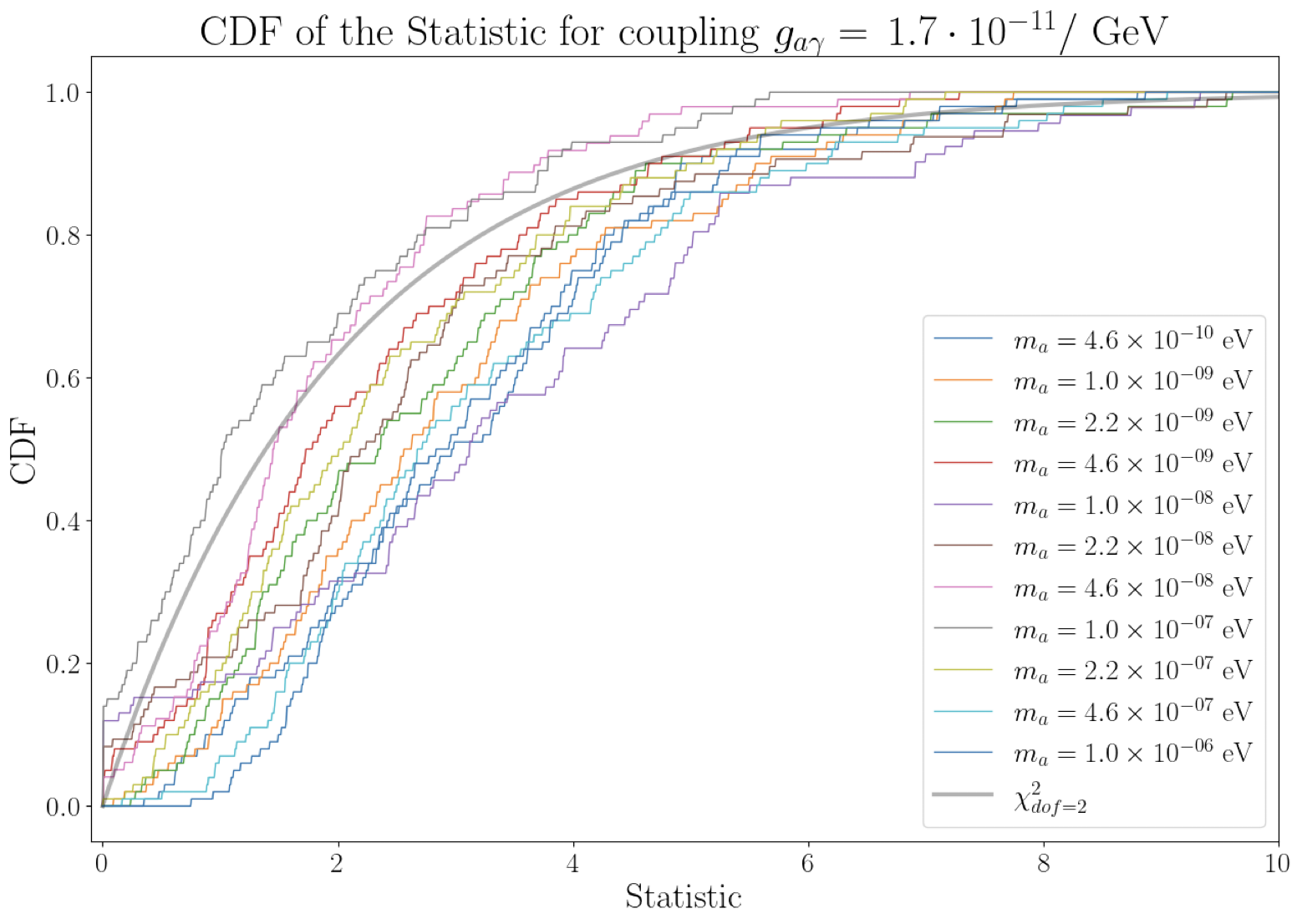}
\caption{
Cumulative Distribution Functions (CDFs) of the statistic $\mathcal{TS}(m_a, g_{a \gamma})$ (see Eq.~\ref{Eq:Likelihood_ratio}) obtained from Monte Carlo simulations for different axion masses ($m_a$) and two axion-photon couplings: $g_{a \gamma} = 5.0 \times 10^{-13}~\textrm{GeV}^{-1}$ (upper plot) and $g_{a \gamma} = 1.7 \times 10^{-11}\textrm{GeV}^{-1}$ (bottom plot). The thicker blue line in the upper plot highlights the CDF for the lowest ALP mass and coupling considered, which aligns with the null hypothesis. The grey line in both plots represents the chi-squared CDF for comparison, illustrating that using the chi-squared CDF with 2 degrees of freedom would lead to undercoverage.
}
    \label{fig:cdf}
\end{figure}

On the upper plot of Fig.~\ref{fig:cdf}, we emphasize (using a thicker line) the CDF for the lowest ALP mass ($m_a$) and coupling ($g_{a \gamma}$) considered in this analysis. Taking into account the telescope's energy resolution, the expected counts under this hypothesis align with those under the null hypothesis (no ALP effect). Indeed both the observed statistic and the CDF obtained from MC simulations are identical for the null hypothesis and for the hypothesis with $m_a = 4.6 \times 10^{-10} $  eV and $ g_{a \gamma} = 5.0 \times 10^{-13}~\textrm{GeV}^{-1}$.

Finally, each distribution of the statistic for each of the 154 ALP points considered is fitted using the gamma distribution $G$:

\begin{equation}
G(x; \alpha, \beta) = \frac{\beta^{\alpha}}{\Gamma(\alpha)} x^{\alpha - 1} e^{-\beta x}.
\end{equation}

Here, $\alpha$ represents the shape parameter, while $\beta$ denotes the rate parameter. The function $\Gamma(x)$ is defined as:
 \begin{equation}
 \Gamma(x) = \int_0^{\infty} t^{x-1}e^{-t} dt.
\end{equation}
The chi-squared distribution with $k$ degrees of freedom is a special case of the gamma distribution $G$, characterized by a shape parameter of $k/2$ and a rate parameter of $1/2$. The fitted gamma distributions are subsequently employed to compute the confidence level (CL) at which each of the 154 ALP hypotheses can be excluded:
\begin{equation}
    \rm{CL} = \int_0^{\mathcal{S}_{obs}} G(x; \alpha, \beta)\:dx ,
\end{equation}
with $\mathcal{TS}_{obs}$ the observed statistic for a given ALP point derived from Eq.~\ref{Eq:Likelihood_ratio}. In order to obtain Fig.~\ref{fig:sigma} each CL is converted to the Guassian equivalent deviation $\sigma$ through the inverse of the error function: 
$
\sigma = \sqrt{2} \rm{erf}^{-1} ( \rm{CL} )
$.

Lastly, it is worth noting that if we had uncritically applied Wilks' theorem and utilized the chi-squared CDF with 2 degrees of freedom (displayed as a reference in grey in Fig.~\ref{fig:cdf}), this would have led to undercoverage. The reason for this is that the chi-squared distribution results in a lower threshold for rejecting a given hypothesis, thereby increasing the likelihood of Type I errors (false positives).

\section{Comparison of spectral counts between null and ALP hypotheses.}
\label{app:counts}

Fig.~\ref{fig:counts} presents a comparison of the observed excess counts per energy bin (multiplied by the center value of the energy bin for visualization purposes) for the three datasets in this work (refer to Tab.~\ref{tab:data}) with those from the null hypothesis model and the best-fit ALP model. As discussed in Sec.~\ref{sec:results}, the latter corresponds to $m_a = 2.15 \times 10^{-8}$ eV and $g_{a \gamma} = 3.81 \times 10^{-12}~\textrm{GeV}^{-1}$. In the figure, one can observe how the expected counts from the ALP hypothesis (blue line) show better agreement with the observed counts (black points) compared to the expected counts assuming the null hypothesis (orange line). Additionally, the flaring state appears to be the most constraining of the three datasets, as it is the only one in which the alternative hypothesis may be significantly favored over the null hypothesis. These facts are emphasized in the bottom part of each of the three plots in Fig.~\ref{fig:counts}, where the relative distance between the observed and expected counts is displayed for all energy bins under both hypotheses, defined as
    $
    ( N_{exc} - s )/ \sigma_{N_{exc}}
$.
In this expression, $N_{exc} = N_{on} - \alpha N_{off}$, $\sigma_{N_{exc}} = \sqrt{N_{on} + \alpha^2 N_{off}}$, and $s$ is given by Eq.~\ref{Eq:EnBin_Integral}.
\begin{figure*}[h!b]
\centering
    \includegraphics[width=0.9\linewidth]{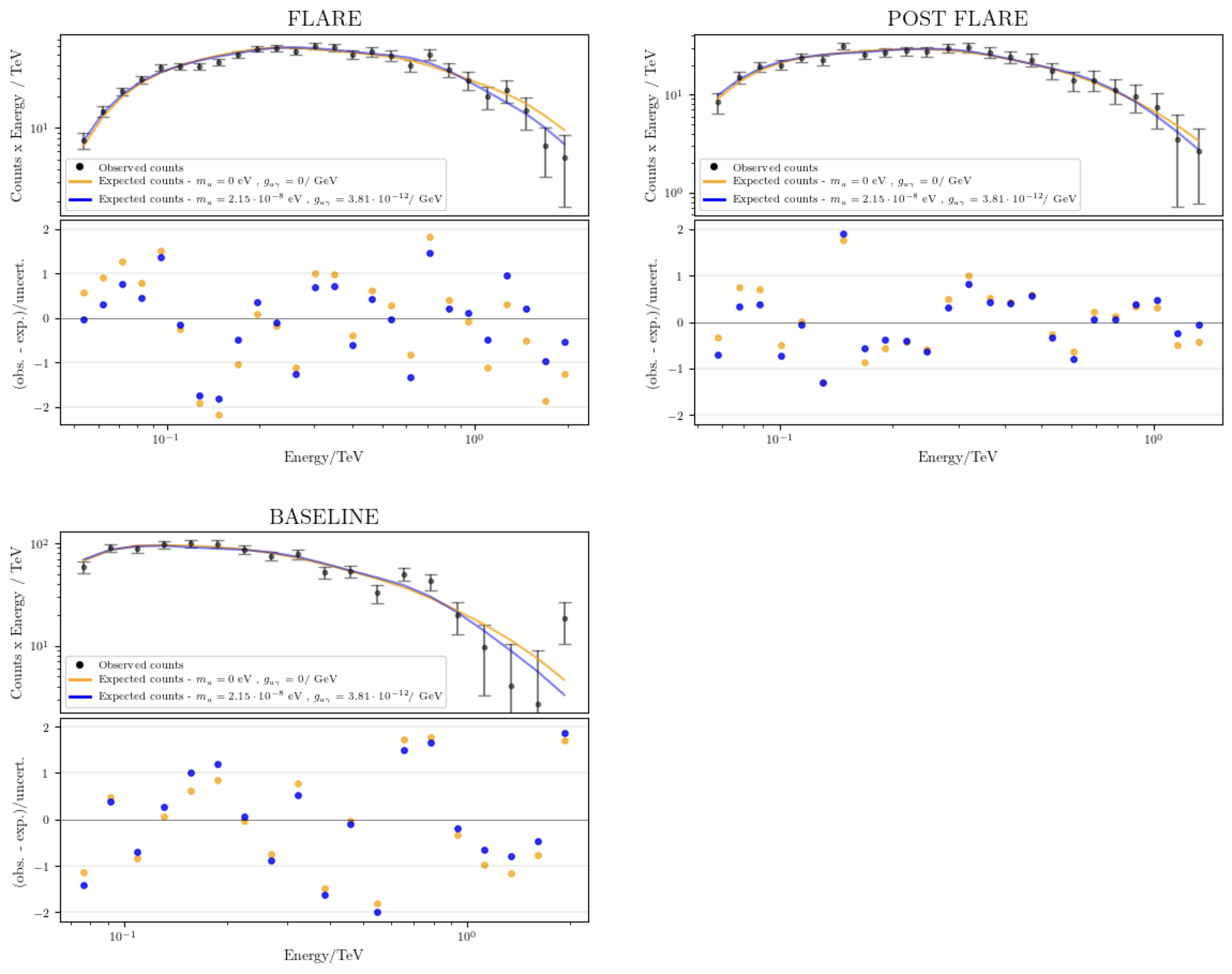}
    \caption{A comparison of observed excess counts per energy bin (multiplied by the center value of the energy bin for visualization purposes)  with those expected from the null hypothesis model (orange line) and the best-fit ALP model (blue line). The expected counts are obtained by applying Eq.~\ref{Eq:EnBin_Integral}, in which the SED parameters are fixed to the values maximizing the likelihood. The observed counts are represented by black points. The bottom part of each plot highlights the relative distance between the observed and expected counts for all energy bins under both hypotheses. This is shown for all the three datasets in  Tab.~\ref{tab:data}, here referred in the title of each plot as ``FLARE'', ``POST FLARE'', and ``BASELINE'', respectively.}
    \label{fig:counts}
\end{figure*}

\end{document}